# The Tribal Theater Model: Social Regulation for Dynamic User Adaptation in Virtual Interactive Environments

Hanzhong Zhang[1], Bowen Duan[1], Haoyang Wang[1], Zhijian Qiao[1], Jibin Yin[1*]

[1] Faculty of Information Engineering and Automation, Kunming University of Science and Technology, Kunming 650221, China

[*] Correspondence should be addressed to Jibin Yin, yjblovelh@aliyun.com

**Abstract**：This paper proposes a social regulation model for dynamic adaptation according to user characteristics in virtual interactive environments, namely the tribal theater model. The model focuses on organizational regulation and builds an interaction scheme with more resilient user performance by improving the subjectivity of the user. This paper discusses the sociological theoretical basis of this model and how it was migrated to an engineering implementation of a virtual interactive environment. The model defines user interactions within a field that are regulated by a matrix through the allocation of resources. To verify the effectiveness of the tribal theater model, we designed an experimental scene using a chatroom as an example. We trained the matrix as an AI model using a temporal transformer and compared it with an interaction field with different levels of control. The experimental results showed that the tribal theater model can improve users' interactive experience, enhance resilient user performance, and effectively complete environmental interaction tasks under rule-based interaction.

**Keywords:** Social regulation; Virtual network; Field; Tribalism

# 1. Introduction

Individual virtual interactive environments can be viewed as inherently structured environments in which people take on designated roles and interact by following well-defined protocols and norms to achieve common specific goals (Almajano et al., 2013). Among these elements, interaction is one of the most important features of virtual interactive environments and is a central part of their conceptual core (Paredes & Martins, 2012). A user's sense of engagement and community in a virtual environment is influenced by many factors. These factors are also major concerns in interactive virtual environments, including the spatial structure of the interaction and the user's avatar.

Place-making, or a "sense of space," is the core technology proposed by researchers in the field of computer-supported collaborative work (CSCW) to promote social interaction in online spaces (Ducheneau & Moore, 2004). In fact, physical space can provide a common understanding of suitable usage and behavior as well as a social explanation of clues in the environment (Amadeo et al., 2009). In the spatial structure of interaction, shared experiences and the collaborative nature of most activities are crucial and, most importantly, integrate an individual into a user community and allow the earning of reputation rewards within it (Hanson et al., 2019).

The design of avatars in virtual environments also affects the interactive experience. Kim (2000) defined an avatar as "an online representation of oneself in a virtual world, aimed at enhancing interaction in the virtual space." Peterson (2005) studied the interaction of non-native users' avatars in virtual environments and concluded that user interaction in virtual environments is beneficial for the development of users' second language abilities, and that the design of avatars can promote online communication and enhance users' sense of participation and community.

At the same time, the development and success of virtual interactive environments are constrained by their usability and sociability. Usability focuses on how users interact to perform tasks, while sociability involves the planning and formulating of social policies to regulate user behavior and interaction within the environment (Preece, 2000). Actually, because virtual interactive environments reflect human evolutionary behavior within them based on user needs and expectations (Preece, 2001), they are dynamic and adaptive in their sociality and usability constraints (Abras et al., 2003).

Therefore, virtual interactive environments can be created to gradually adopt new types of interaction rules rather than promoting specific interactions. One of these new methods is interaction regulation, which refers to how cooperation in work can occur in a non-entangled way by delineating and defining communication rules. Becker-Beck et al. (2005) argued that the reference level of rules is defined as "conceptual relationships," which refers to the process of letting a group achieve common cognitive representations. Therefore, interaction regulation includes individual regulation that defines individual interactions and behavioral rules and organizational regulation that focuses on the coordination of individual interaction and how to organize within the environment.

There are many existing studies on individual regulation. Brota et al. (2009) proposed a generic interaction framework, iObjectDoor, that allows the creation of avatars conforming to the norms of a multi-agent system to provide different object visualizations depending on the client, and to manage behaviors in the environment. In a dynamic space, the participants are then informed about the evolution of the activity, and norms are used in it to organize their actions to define the consequences of the actions and to prevent undesired participant behavior.

However, the practice of using limiting rules of individual regulation with users also affects the availability of virtual environments. For example, Almajano et al. (2013) proposed a framework for hybrid regulation using virtual institutions (VIs) to design a hybrid and structured virtual environment. However, the inherent limitations of rule-based interface dialogs (e.g., as they are implemented in command-based robotic dialogs) make some users uncomfortable.

Paredes and Martins (2012) proposed a model to create a canonical interaction environment based on a theater metaphor that dynamically adapted interaction rules to meet user needs. This model is typical of an interaction rule-based approach in virtual interactive environments. Based on the authors' constructed architecture named ASTeaS, it obtained better interaction results. However, the model also failed to go beyond the limitations of defining interaction rules on user subjects. Because users were categorized in advance to assign system roles and interfaces that corresponded to them, the model cannot actually adapt dynamically to interaction rules at each moment in time and provides users with a sense of excessive digital surveillance.

Akar and Mardikyan (2018) derived the existence of different user types in communities and their motivations by collecting and analyzing user behavioral data and successfully managing these communities to provide better services. However, this study overlooked that the members of these communities are likely changing, so user types could also be affected, which would impact the self-adaption.

Indeed, the limitation of individual regulation that controls a user's behavior itself is that in emphasizing the importance of the task, it sacrifices a user's operational experience. This leads directly to the neglect of user subjectivity in such research.

One approach involves resilient performance, which refers to the key functional attributes of complex social technology systems that can prevent system crashes in the face of expected and unexpected changes and disturbances. It is generated by the self-organized improvisation and decision-making of individuals. Disconzi and Saurin (2022) point out that previous studies that have referred to the concept of design for resilience (DfR) have often neglected the role of humans (Uday & Marais, 2015) in adopting an overly technical perspective (Kusiak, 2020). Thus, Disconzi and Saurin (2022) proposed the concept and principles of design for resilient performance (DfRP), which emphasize the role of human factors in system design, including the support of continuous learning at the individual and organizational levels. Their proposal of the DfRP actually fully considers human subjectivity, emphasizing the role that humans play in the studied system. Many further studies have demonstrated that utilizing user subjectivity can lead to better interaction results.

For example, Benlamine et al. (2021) studied a framework for game element adaptive player emotional states that changed the layout of the game elements in the interface by capturing the facial expressions of users during interactions. Their research results indicated that adapting games to users' personal feelings can enhance their experiences.

Bezerra and Hirata (2011) studied individual and environmental factors that influence motivation in virtual communities and explored the mechanisms that increase motivation in virtual communities. Increasing the enthusiasm of virtual community members indirectly promoted their participation.

Krishna et al. (2019) established a role-based regulation model that estimated learners' regulation states through their behaviors, cognition, and motivation to improve their self-regulation skills by implementing adaptive regulation strategies as needed. The disadvantage of this model was that the role of the framework was limited and adaptive, based only on the existing roles.

However, both the user experience and attention to tasks cannot be ignored in virtual network environments. The practice of tilting resources in favor of the user simply by diminishing individual regulation or by some other mechanism certainly increases user subjectivity, but it also naturally decreases the tilting of resources in favor of the task. Therefore, the importance of organizational regulation needs to be recognized and explored. For example, a more matriarchal model of group consciousness could be a powerful tool. Strauss and Rummel (2021) confirmed the role of group consciousness tools. Their research results indicated that timely adjustment of user interaction processes based on user behavior sequences can help groups achieve equal participation.

Therefore, an interaction model that focuses on organizational regulation without suppressing user subjectivity and with more resilient user performance needs to be discussed. Given the superiority demonstrated by Paredes' social theater model (based on a theater metaphor), it is feasible to construct the interaction model we need by combining it with an appropriate sociological model.

In fact, the literature in sociology does not appear to provide us directly with the specific mature theoretical tools we need for this particular type of application. Therefore, to enhance user resilience performance, or in other words, to address the core issue of "enhancing user interaction freedom" in virtual interactive environments, this article proposes a social regulation model for users' dynamic adaptation in virtual interactive environments based on Bourdieu's field theory and Maffesoli's tribalism, namely the tribal theater model (TTM). This model emphasizes the subjectivity of the user; that is, the model focuses on how to maximize the user's personal interactive experience through organizational regulation based on deep learning algorithms while minimizing the need for monitored individual regulation without letting the interaction fall into unconstrained chaos. Compared to the traditional social theater model used for virtual network interaction models (Paredes & Martins, 2012), the interaction models constructed through the TTM are expected to have greater user resilience performance.

The proposed interaction model should satisfy the following hypotheses:

- H1  In a virtual social environment, the TTM will provide better user interactive experience and enhance user resilience performance.

- H2  In a virtual social environment, the TTM will effectively complete the environmental interaction tasks that can be completed under rule interaction.

Among them, H1 will be used to determine the advantages of the TTM relative to the social theater model in emphasizing subjectivity, and H2 will be used to ensure the effectiveness of the TTM's relatively unconstrained interaction. In Section 4, these two hypotheses are validated separately.

Section 2 of this paper introduces the sociological aspects of the TTM; Section 3 introduces the TTM in human–computer interaction using a chatroom as an example, as it is a dynamically regulated social model centered on an artificial intelligence matrix; Section 4 introduces the experimental evaluation of the TTM and discusses the results obtained; and Section 5 presents some conclusions and discusses future work.

## 2. Sociological Analysis of the TTM

Before formally proposing the TTM, some core concepts in the sociological background need to be described. Readers who are more interested in TTM as a human–computer interaction model can skip this section and proceed to Section 3 directly. The process of formally proposing the TTM follows the progressive process shown in Figure 1.

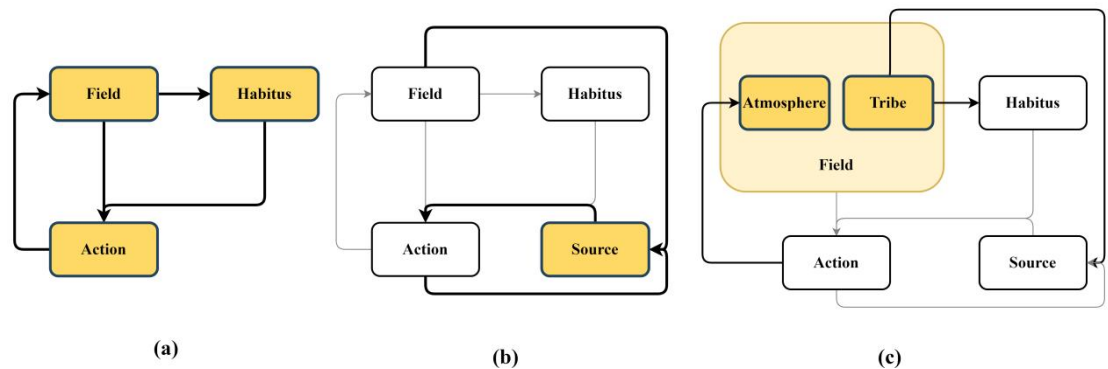

**Figure 1.** The construction process of the TTM. (a) The basic structure of the TTM. (b) The overall structure of the TTM. (c) The detailed structure of the TTM.

The core concepts of the tribal theater model are naturally the "tribe" and "theater." The sociological concept of a "tribe" was refined primarily by Maffesoli (2022). He pointed out that a tribe refers to a kind of community. People naturally gather together for some reason to form social groups. Furthermore, he connected the concept of the "tribe" with that of a "theater." Maffesoli (2022) pointed out that "the characteristic of social relations is that individuals can have functions in society…. As individuals change stage costumes based on their (cultural, religious, friendship) taste, they will gain a place in various games of the 'World Theater' every day" (p. 104). It can be seen that Maffesoli's "tribe" is part of the "theater" in this description.

The theater is actually a metaphor that is highly consistent with Bourdieu and Wacquant's (2015) description of a "field." A field can be understood as a place of interaction. In their hypothesis, each field prescribes its own peculiar regulating principle. These principles define an interaction space. In such a space, actors compete based on their position in the space to change or attempt to maintain the scope or form of their space.

Merleau-Ponty (1983) provided a case in point. When playing rugby, the field is seen by the players as covered with various constraint lines that divide and connect, for example, different kicking areas that require players to take certain determined actions. Each action taken by the players adjusts the characteristics of the field in their eyes and establishes new offensive routes. Their actions, in turn, unfold within a new range, once again changing the field as it is perceived by the players (Merleau-Ponty, 1983).

It can be seen that the field is changed by the actions of the users, but what is directly changed is not the "tribe" aspect of the field mentioned earlier, but another more abstract aspect, namely its "atmosphere" (Norberg-Schulz, 2010). In turn, the content of the action is strongly related to the current field. However, it is obvious that each person may take different actions when facing the same scene. Therefore, the generation of action also requires an individually related factor, namely, the "habitus." This concept is similar to what we would understand as the habits of an individual (it is actually the Latin word for "habits," and its plural form is "habiti"), but in the TTM, it can be understood as a decision tree of actions. When we are within an environment, a corresponding decision tree is activated, and we decide on an action to take based on our action habits. This is similar to the logic of reinforcement learning, in which the environment and agents influence each other. The field shapes the habitus, and the habitus becomes a product manifested by the body as a result of the inherent and inevitable attributes of a certain field (Bourdieu, 2017).

From this, we can obtain a basic closed-loop structure, such as that shown in Figure 1(a): the field shapes the habitus, and the field and habitus together lead to action, which in turn changes the field. We can continue the analogy with the logic of reinforcement learning: in reinforcement learning, feedback from the environment constructs the model of the agent, the environment and the model together lead to the action of the agent, and the action of the agent leads to changes in the environment.

In fact, if we had only a structure similar to reinforcement learning, it would still be insufficient to model human interactions (Hackel et al., 2020). According to Grenfell (2012), Bourdieu thus pointed out a key to interactive regulation, namely "capital." Sociology uses this term in a broad sense to expand the concept and enable it to be applied to complex interpersonal networks and to different types of transformations and exchanges between different fields. Concepts such as literacy level and social status have been introduced into the scope of capital and play a role in interaction scenes. In the TTM, it can therefore be referred to as "resources."

For example, when we play a certain game, the various strategies of players and the different factors that determine their game are not only a function of the quantity and structure of their resources at the moment of consideration but are also a function of the evolution of the quantity of these resources and their structure over time, that is, a function of their social trajectory.

Therefore, changing the distribution and relative weights of the various resource forms is equivalent to changing the structure of this field. By considering fields in this way, they all have a certain degree of dynamic change and adaptability and avoid the inflexibility and resilience of social theater (Lehmann, 2006) or traditional structuralism, such as reinforcement learning, which was mentioned earlier.

On this basis, Bourdieu and Nice (1980) provided a formula for action (with some modifications):

$$Action = f(habitus + resource) + field \qquad (1)$$

As described in Figure 1(b), on the basis of the original closed loop, resources participate in the generation of action, while the resources themselves are determined by the combination of the field content and action. Figure 1(c) provides a more detailed division; that is, actions affect the atmosphere portion of the field, relationships and resources are influenced by the tribe portion of the field, and the overall field (including the tribe and atmosphere) directly leads to actions as an environment. The role of this more detailed division will be provided in the next section.

From this equation, we can obtain the TTM as a foundational sociological model. As mentioned above, the structure of this model provides flexibility and adaptability that is not contained in traditional models of social theater. In social theater and a series of similar theater metaphor approaches, actors must perform with scripts; this provides an order but loses the subjectivity of the actors themselves (Nacache, 2007). However, the TTM operates in the style of Boal (2008). There is no longer the need for an order that provides requirements, serves as a "director," and forces individuals to follow. In contrast, people have gathered into a small-group consciousness, that is, in the matrix of a tribe; it has taken the position of the director. According to Simmel (2002), this is explained by Zimmer and his sensory sociology as a stage of "shared by all."

## 3. Structure of the TTM

In Section 2, we connected several important concepts and proposed a tribal theater model that is more resilient than traditional social theater models. As depicted in Figure 1, the conceptual chain of the TTM is connected head to tail as a closed loop, which constitutes a de-identified interaction structure.

Based on this sociological model, we can establish its relationship with a virtual social network interaction scene; that is, the interaction environment of social networks serves as the field (in which the interaction space serves as the tribe, and the user interaction theme serves as the atmosphere), the user's operating habits serve as the habitus, the materials required for interaction serve as the resources, and a model for resource allocation serves as the matrix.

Public spaces such as chatrooms are typical complex social technology systems with characteristics such as diversity, dynamism, and uncertainty, which can be used as a typical example to verify user resilience performance. As an example, and to provide definitions of the corresponding concepts in the virtual social network interaction model, we consider designing a chatroom environment as an instantiation of the TTM.

### 3.1 Field

In the sociological definition of the TTM, the structure of the field can be considered composed of the atmosphere and the tribe and analyzed based on the classification of "space" and "characteristics." "Space" implies the elements that constitute a field, that is, the tribe. "Characteristics" generally refer to the atmosphere, which is the most abundant attribute in any place.

Therefore, in a virtual network environment, a number of key terms can be defined. A tribe is a social space in which all interactions are limited within a certain group. The atmosphere is a theme of interaction, and all interactions revolve around the atmosphere. Specifically, in the example of a chatroom, the field of an online chatroom is generated by a chat group provided by a platform and a chat topic. This topic can be proposed by chatroom members, or it can also be generated based on an order when everyone agrees on the need to complete a task, whether voluntary or involuntary (just as, no matter how unwilling, students always have to gather together in groups to organize and create a classroom presentation).

## 3.2 Habitus

The habitus is not only a physical manifestation but also a psychological and social manifestation. It includes a range of ways individuals apply perception, cognition, and action (Bourdieu, 2017).

Therefore, in the TTM, habitus can be considered a series of user perceptions and actions applied in interacting within a specific field, which then comprise a decision tree of user action. According to different field conditions, users exist and are shaped by different habiti. For example, in an online chatroom, users may develop different habiti (such as what language to use, what emotions to express, and what identity to display) based on factors, such as chat group members, topics, tasks, and orders. These habiti not only reflect the user's personality and preferences, but also reflect their cognition and evaluation of this online chatroom.

## 3.3 Resources

In the TTM, the materials relied on by user action can be seen as resources. Users have their own initial resources, which can be used to carry out their own actions and acquire resources from such actions. In the production of such resources, their allocation is carried out by the field under the control of the matrix. Referring to the chatroom example, a resource can be a voting mechanism within a chatroom. In the voting process for selecting temporary chatroom administrators, users initially have their own votes for the election. When a user becomes a temporary administrator of the chatroom through the voting mechanism, this vote will not only be replenished, but the user will also gain more votes in subsequent voting activities due to the benefits generated by previous voting actions (such as becoming an administrator).

## 3.4 Action

Action is generated by a combination of field, resources, and habitus. Such an action will, in turn, affect resources and the field (the reason why the habitus is not changed by action, as we mentioned in Section 2, is that it is indirectly changed through the field). Just as in a virtual environment of a chatroom, after we have had a chat, the topic in the chatroom field will change due to the change of topic. As the initiator of the topic, the user's resources (such as prestige) will increase with it.

## 3.5 Matrix

The concept of the matrix is the core of the instantiated TTM. Due to the de-identification of the model and the precariousness of the virtual field, to avoid a chaotic interaction, there needs to be a symbol of order that replaces rules and the collective subconscious for regulation, that is, what Bourdieu (1987) describes as "a generative matrix that exists as a social variable". At the same time, resources also require a mechanism for circulation and allocation and an executing entity. Obviously, the field of the virtual space does not have the premise of Bourdieu's (1987) statement that the matrix is "constructed in history and rooted in institutions", so we need a more specific alternative.

Due to the premise of the de-identification of the TTM, it is not possible for a human to act as a transcendent matrix (such as chatroom administrators in traditional structures). As an alternative, a transcendental and proxy-controlled algorithm model can be considered.

This model should follow a structure that takes as joint input the action vector and the field vector quantified by the field content and as output the vector of new resource structures assigned after the action. Its structure is shown in Figure 2. One possible implementation is deep learning technology. The form of vector input and output in deep learning, as well as its retraining mechanism, make it very suitable for completing such tasks.

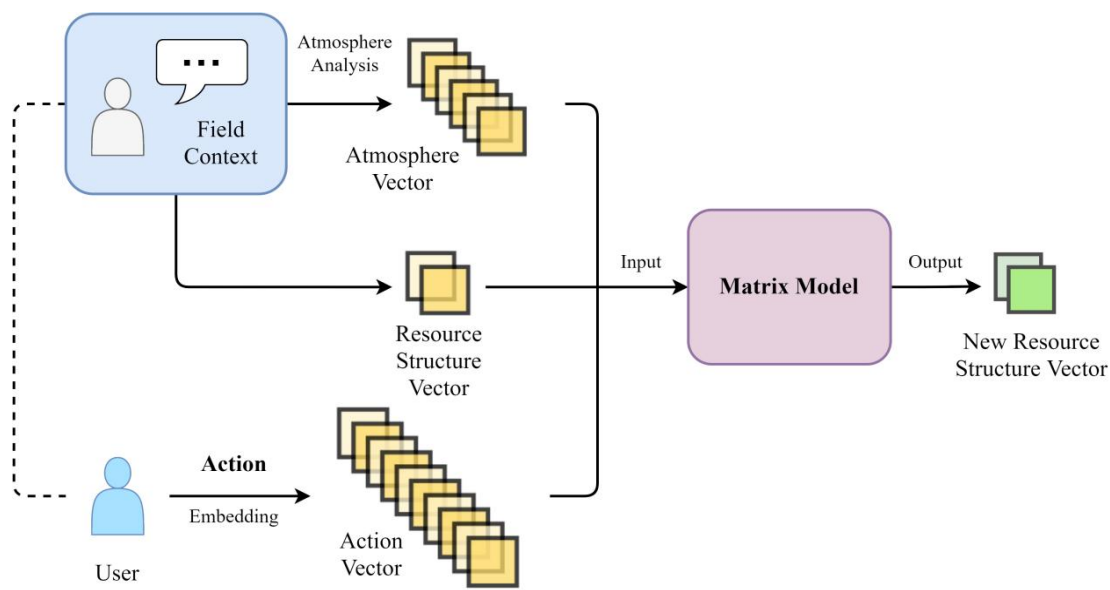

**Figure 2.** The matrix model of the TTM.

Therefore, we obtained a tribal theater model for social regulation and user dynamic adaptation in a virtual network environment.

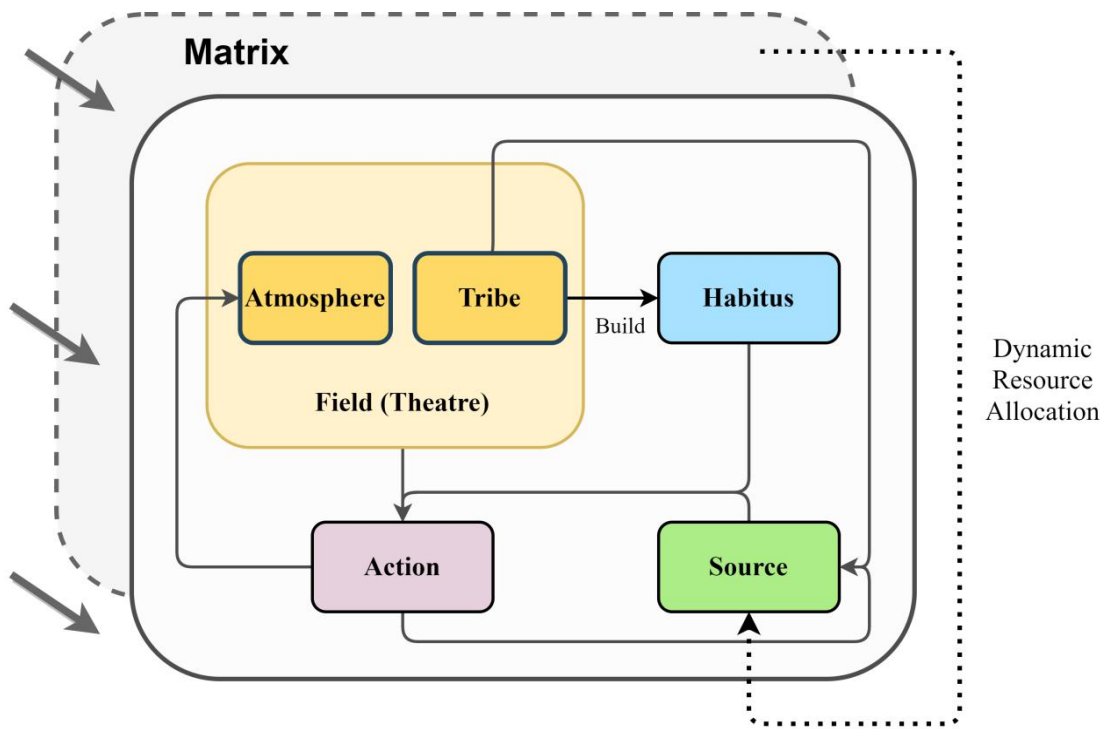

**Figure 3.** Structure of the TTM.

## 3.6 Interaction Flow: Taking a Chatroom as an Example

In the other parts of Section 3, we considered the definition of the core concept of the TTM in the chatroom instance. From this, we can consider the entire interaction flow of the TTM in the chatroom case, as shown in Figure 4.

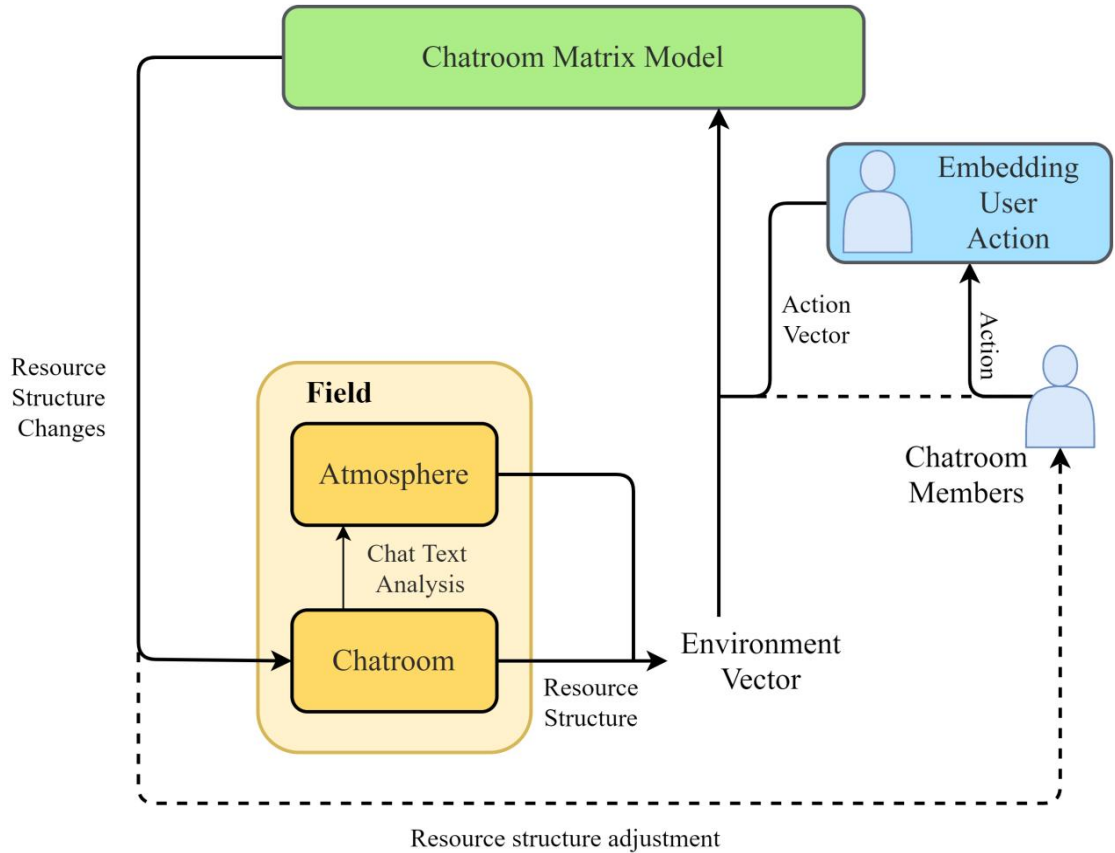

**Figure 4.** The interaction flow of the TTM in the chatroom.

In a chatroom based on the TTM, there is no identity designated as an administrator. Each member has a certain amount of resources. The matrix model provides the most basic action options, such as speaking, withdrawing speech, issuing tasks, and voting. Executing an action within it will consume a certain amount of resources and then delegate the action to the matrix model for processing. Along with an action, the model takes as input a vector representing the current resource structure of the chatroom and a quantified atmosphere vector derived from a text analysis of the chatroom's messages. Based on this input, the matrix will adjust the resource structure within the chatroom and output it as a vector.

Therefore, we can obtain the interaction flow of the chatroom instance based on the TTM. This is shown in Algorithm 1.

---

**Algorithm 1.** Interaction flow of the chatroom instances.

---

Initialize the user behavior embedding model UM  
Initialize the model for analyzing chatroom atmosphere SKEP  
Initialize the matrix model of chatroom $Matrix$  
Initialize the user action α  
Initialize the resource structure vector of chatrooms $v_r$  
**Repeat** (for each action):  
    Read α, $v_r$  
    $v_{\text{atm}} \leftarrow \text{SKEP}(v_r)$  
    $v_e \leftarrow v_r + v_{\text{atm}}$

---

$v_a \leftarrow \text{UM}(\alpha)$

$v_r' \leftarrow Matrix(v_a, v_e)$

Update resource structure with $v_r'$

## 4. Experiments

The implementation of the TTM and the experimental method used to examine it were based on case study scenes. To test the hypotheses proposed in Section 1, we evaluated the user interactive experience, user resilience performance, and completion of environmental interaction tasks to verify the effectiveness of the TTM. Among them, the user interactive experience was evaluated through an interactive quality indicator, while the user resilience performance was evaluated through an interactive freedom indicator.

To conduct experiments based on the de-identification of the field, we created a controlled evaluation environment using the social software Tencent QQ. For comparison purposes, we also collected data in a traditional presence identity environment without any regulation or matrix.

## 4.1 Experimental Presupposition

### 4.1.1 Experimental Participants and Environment

The main purpose of the experiment was to verify the hypotheses in Section 1, that is, whether the TTM was suitable for providing a better user interactive experience and enhancing user resilience performance, and whether it could help users complete environmental interaction tasks complete under rule-based interaction. To achieve this, an experimental scene had to be constructed.

It is generally known that there is a positive correlation between user interaction enthusiasm and a familiar environment. Thus, constructing an experimental environment that was unfamiliar to the participants would not be conducive to examining the factors that impact the effectiveness of the model. Therefore, we decided to rely on the existing social software Tencent QQ to complete the construction of the experimental scene and used a chatbot created specifically for the experiment.

Tencent QQ is a popular WhatsApp-type online chatroom in China. Currently, the vast majority of online socializing among Chinese teenagers takes place on it. There is a hierarchical distinction among users who participate in one of its chatroom groups. The group master, as the creator of the group and the possessor of the highest permissions, has the power to designate group chat rules and punish those who do not comply. Administrators can be designated by the group master, with lower levels of permission than the group master, but these individuals can also regulate the speech standards and topics of group members. By convention, the group master and administrators also have an obligation to direct the trends of topics within the group. Thus, the structure of QQ itself provides very strong ordered interactions.

In order to build a comparative experimental environment, we required such a setting. Therefore, we created a chatbot using the communication protocol of Tencent QQ. It could collect chat messages within a group, and when used as a virtual group master, it could process chat resource allocation through mechanisms such as "mute," "set administrator," "remove members," "withdraw message," and "speech frequency limit," which were associated with any group.

To evaluate the effectiveness of the de-identified TTM, we set up an experimental environment with three chat groups on QQ. Two of these were established based on an existing group, and one was newly created to be adapted to the TTM. The members of the three groups had basic knowledge of each other. The key participants in all three groups were informed about the experiment.

The descriptions of the experimental groups are as follows:

- *Group 1 (high control).* This was a group with a primary active population of 20–30 people. The group had a clear hierarchy, and the administrator, as the manager of the group, basically controlled the direction of the group's chat topics and removed group members who violated the rules. The chatbot was here only to collect and analyze chat content and did not participate in the resource allocation of the chat field.
- *Group 2 (low control).* This was a group with a primary active population of 20–30 people. There was no hierarchy in the group. The group master, who was the creator of the group, was not involved in the discussion and did not set up any administrators, so the group members spontaneously chatted with each other, and there were often situations where multiple topics coexisted. The chatbot was here only to collect and analyze chat content and did not participate in the resource allocation of the chat field.
- *Group 3 (main field).* This was a group of 10 people created at the beginning of the experiment and was used only for the experiment. The group master was the experimental chatbot, and there was no administrator. The group structure followed the TTM, and the chat resources were distributed by the chatbot.

### 4.1.2 Experimental Evaluation Tools

Based on an existing questionnaire of user satisfaction with a human–computer interface (Chin et al., 1988), we created a customized questionnaire for evaluating the environment, divided into several parts, reflecting each parameter in the evaluation. Each parameter was evaluated by the users based on a scale between 0 and 9. These parameters included user interactive quality, user interactive freedom, and digital surveillance sensation.

To quantify the atmosphere and field changes, we used the sentiment knowledge enhanced pre-training (SKEP) model (Tian et al., 2020) to classify the emotional tendencies of chat texts and assign the proportion of positive and negative emotional tendencies to samples. Specifically, this model assigned the chat text a positive probability $P$, a negative probability $N$, and a confidence level $C$. These three values were all within the range of [0, 1]. To simplify the calculation, we defined an atmosphere value for a chat text using the following formula:

$$Atmosphere = (P - N) * C \qquad (2)$$

As a result, the atmosphere value of a chat text will be within the range of [−1, 1]: less than zero is a negative atmosphere, and more than zero is a positive atmosphere.

### 4.1.3 The Matrix Model

To allocate resources, we needed to train a matrix model. The structure of the matrix model is detailed in Section 3.5. In this study, we used a temporal transformer model to train the matrix model. This is a time-series neural network structure based on the transformer architecture (Vaswani et al., 2017), which can effectively deal with long-term dependencies and multivariate time series. The core idea of the temporal transformer is to use a self-attention mechanism to capture local and global dependencies in temporal data while using positional encoding to preserve temporal information. A temporal transformer can be divided into two parts: the *encoder* is responsible for translating the data into high-dimensional feature vectors, and the *decoder* is responsible for generating a prediction of the current state based on the output of the encoder and the historical state.

To facilitate the engineering implementation, we also defined the parts of the matrix model in this experiment. Resources are defined here as the number of texts that can be sent during a certain period of time. The resource structure vector, as an initial input, contains the number of resources (1-dimensional) and the proportion of resources (1-dimensional) of the actor at this particular moment in time. The atmosphere vector (10-dimensional) is a vector composed of the textual analysis of the first 10 messages of an action at a particular moment, each element of which is an interval of value from −1 to 1 (from negative to positive). The action vector (1,024-dimensional) is a vector of sentences produced by a user at a particular point in time, which was obtained by partitioning the sentences into words, assigning weights to the resulting words using the term frequency–inverse document frequency (TF–IDF) algorithm, and then weighted averaging the word vectors of these words. As the output of the new resource structure vector, for the convenience of implementation, we set the output directly as a new value of the current actor's resources to achieve the effect of changing the global resource proportions. The engineering implementation of the matrix model is shown in Figure 5.

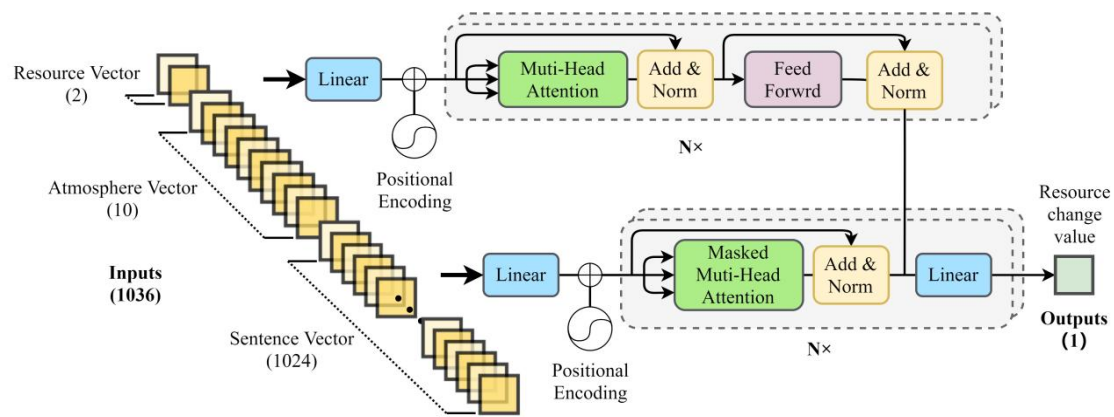

**Figure 5.** Engineering structure of the matrix model.

To train this AI matrix, we first established a dataset. We chose a group that interacted according to specific rules. A resource quantity of five messages per minute was first set and then dynamically labeled according to the administrators' subjective assessments. The administrators increased or decreased the resources of users through the speech restriction and banning functions. The collected dataset contained a total of 7,936 valid data, and all the data were processed according to the above method; that is, each individual data point contained 1,036 dimensions.

In this paper, we used the PyTorch framework to implement a temporal transformer model, where the encoder input dimension was 2,048, the number of multi-attention heads was 8, the activation function was a rectified linear unit (ReLU), the number of sub-encoder layers was 6, the dropout was 0.1, the number of sub-decoder layers was 6, and the epsilon of the layer normalization was 1e−5. The model used mean square error (MSE) as the loss function, the stochastic gradient descent (SGD) optimizer was used to optimize the parameters of the model, the learning rate was set to 1e−5, and the early stopping method was used during training to prevent overfitting. When the value of the loss function on the test set did not decrease for 10 consecutive epochs, the training was stopped, and the optimal model parameters were saved.

After obtaining the matrix model, we asked 10 people who did not participate in subsequent experiments to sample and evaluate the model on the test set. The score of the obtained model was 3.50 ± 0.04 (out of 5).

## 4.2 Experimental Analysis and Results

We collected chat data from Groups 1 and 2, and required members of Group 3 to transfer their chats from other places to this group as much as possible within a week. As part of this process, we organized a thematic discussion in each of the three groups.

After the experiment, a questionnaire was administered to the key participants involved. A total of 62 questionnaires were collected.

From Group 1, we collected a total of 27 questionnaires, including from 18 males (66.7%) and 9 females (33.3%); 11 people aged 17 and below (40.7%), 11 people aged 18–22 (40.7%), 3 people aged 22–25 (11.1%), and 2 people aged 25 and above (7.4%); 11 people (40.7%) had a secondary education, 10 people (37.0%) had a bachelor's degree, 3 people (11.1%) had a master's degree, and 3 people (11.1%) were already working; 12 people (44.4%) majored in the humanities and social sciences, 13 people (48.1%) majored in the natural sciences, and 2 people (7.4%) majored in other fields (including law and medicine).

From Group 2, we collected a total of 25 questionnaires, including 19 males (76.0%) and 6 females (24.0%); 8 people aged 17 and below (32.0%), 13 people aged 18–22 (52.0%), 3 people aged 22–25 (12.0%), and 1 person aged 25 and above (4.0%); 7 people (28.0%) had a secondary education, 12 people (48.0%) had a bachelor's degree, 2 people (8.0%) had a master's degree, and 4 people (16.0%) were already working; 10 people (40.0%) majored in the humanities and social sciences, 11 people (44.0%) majored in the natural sciences, and 4 people (16.0%) majored in other fields (including law and medicine).

From Group 3, we collected a total of 10 questionnaires, including 6 males (60.0%) and 4 females (40.0%); 5 people aged 18–22 (50.0%), 5 people aged 22–25 (50.0%); 4 people had a bachelor's degree (40.0%), and 6 people had a master's degree (60.0%); 4 people (40.0%) majored in the humanities and social sciences, 4 people (40.0%) majored in the natural sciences, and 2 people (20.0%) majored in other fields (including law and medicine).

An analysis of variance (ANOVA) was conducted on the questionnaire, revealing significant differences in the mean values among the groups. For the control level, $F(2, 59) = 39.53$, with $p < .01$. For interactive quality, $F(2, 59) = 4.14$, with $p < .05$. For interactive freedom, $F(2, 59) = 7.09$, with $p < .01$. For digital surveillance sensation, $F(2, 59) = 3.68$, with $p < .05$. For task completion results evaluation, $F(2, 59) = 3.32$, with $p < .05$. These results indicate at least one significant difference in the comparisons of mean values among the groups.

The mean comparison results indicated that Group 3 had significantly higher scores for interactive quality (7.30±0.54) and interactive freedom (7.70±0.37) than Group 1 (5.85±0.35 and 5.78±0.35, respectively) and Group 2 (7.08±0.35 and 7.00±0.27, respectively), which is consistent with the prediction of H1. However, on the digital surveillance sensation (reverse scored), Group 3 (3.80±0.85) scored significantly lower than Group 1 (3.93±0.36) and significantly higher than Group 2 (2.48±0.37).

On the task completion results evaluation, the Group 3 score (7.20±0.39) was significantly higher than that of Group 1 (5.74 ± 0.45) and Group 2 (6.80 ± 0.27), which was consistent with the prediction of H2. A summary of the questionnaire analysis data is shown in Table 1, and the between-group mean comparison results of the questionnaires are shown in Figure 6.

**Table 1.** Summary of questionnaire analysis data.

|  | Group 1 (High Control) | Group 2 (Low Control) | Group 3 (Main Field) |
|---|---|---|---|
| **Control Level** | 6.22±0.21 | 2.36±0.28 | 4.90±0.60 |
| **Interactive Quality** | 5.85±0.35 | 7.08±0.35 | 7.30±0.54 |
| **Interactive Freedom** | 5.78±0.35 | 7.00±0.27 | 7.70±0.37 |
| **Digital Surveillance Sensation*** | 3.93±0.36 | 2.48±0.37 | 3.80±0.85 |
| **Task Completion Results Evaluation** | 5.74±0.45 | 6.80±0.27 | 7.20±0.39 |

*Reverse scored

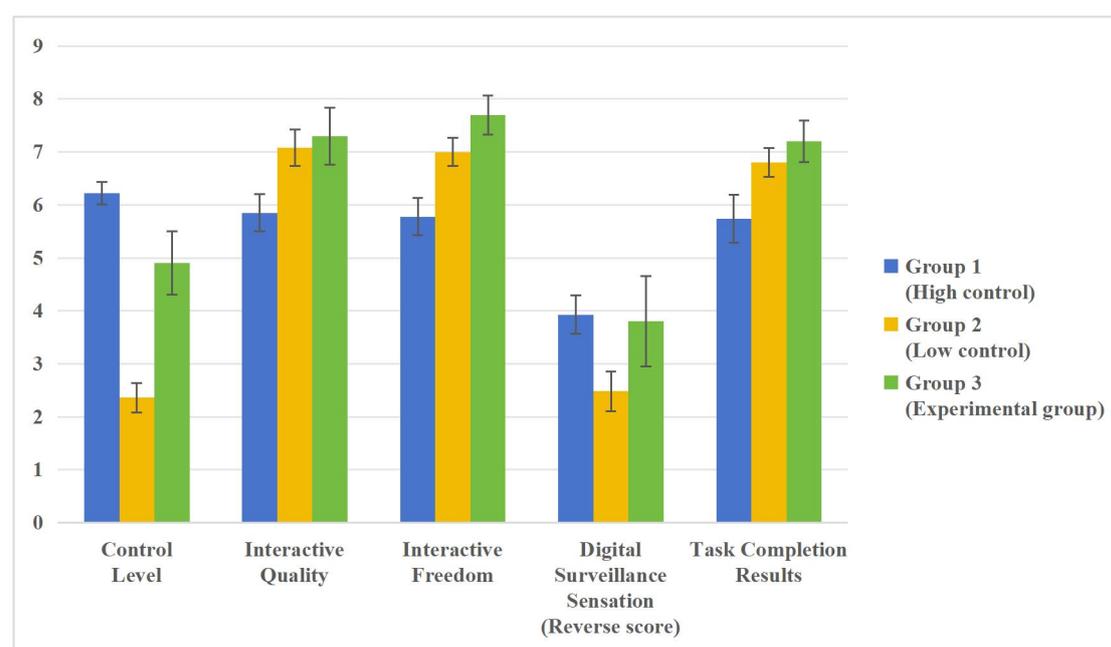

**Fig. 6.** Between-group mean comparison results of the questionnaires.

We also randomly sampled chat messages from Groups 1 and 2 for a week, as well as chat messages from Group 3 during the experiment. We collected 400 points of chat text data from each of them, and used the SKEP pre-training model to classify their respective natural language emotions, assigning emotional tendency labels to all the samples for a field atmosphere analysis using the formula described in Section 4.1.3.

The atmosphere analysis results showed that the score of Group 3 (−0.05 ± 0.04) was higher than that of Group 1 (−0.19 ± 0.03) and slightly lower than that of Group 2 (0.05 ± 0.04). This further supports the premise that the TTM can maximize the user's personal interaction experience while ensuring orderly interaction.

## 4.3 Discussion

Overall, the experimental results supported H1 and H2, that is, the TTM can better improve the user interactive experience, enhance user resilience performance, and help users complete environmental interaction tasks under regular interaction. However, there were some results that required further discussion and analysis.

In terms of the digital surveillance sensation, the main field (Group 3) scored lower than the high control group (Group 1) and higher than the low control group (Group 2). This indicates that the TTM still provided users with a digital surveillance sensation to a certain extent, which limited their freedom of interaction in the environment. One possible reason for this was the limitations imposed by using QQ as an experimental field. Indeed, a familiar interaction environment can provide users with a more natural and daily interactive experience and lead to more accurate interaction behavior. However, the limitations of the QQ environment determined that when the resource quantity was too low, the restriction on users' speech could only be achieved by muting them, which inevitably communicated to users the feeling of being supervised. Therefore, we can say that, based on the results obtained, we cannot conclude whether the TTM necessarily increases a digital surveillance sensation.

The atmosphere analysis results from the chat messages retrieved from the three experimental fields also contained an important finding. With a higher digital surveillance sensation score than the comparison field, the main field still had a higher chat content emotional tendency score than the heavily regulated high-control group. This not only confirmed the previously analyzed questionnaire results that the TTM performed relatively better in experience indicators, such as interactive quality, but also implied that there may be other factors that lead to the observed increase in the digital surveillance sensation.

In summary, the TTM improved the user interactive experience, enhanced user resilience performance, and helped users accomplish interaction tasks. This indicates that the TTM was effective and has the potential to become an important tool for future digital interaction design. However, there are still some issues. The user still perceived a digital surveillance sensation, which may be due to limitations in the experimental field environment.

## 5. Conclusion

In this paper, we proposed and discussed a social regulation model for dynamic user adaptation in a virtual interactive environment, namely the tribal theater model (TTM). The goal of this paper was to establish an interaction model that focuses on organizational regulation without suppressing user subjectivity and that has more user resilience performance. Sociological analysis was used to establish the theoretical foundation of this model and migrate it to the engineering implementation of virtual interactive environments. The model defines user interactions within the field, which are regulated by a matrix model through resource allocation. We evaluated the TTM based on research of chatroom cases. The results indicate that the TTM can improve the user interactive experience, enhance user resilience performance, and help users complete interactive tasks.

Future work should optimize the regulation mode of this model, restrict the interaction of low-resource users, encourage the interaction of high-resource users in a more-gentle way, and reduce the perceived presence of the regulation matrix. In terms of the dynamic adaptation of the system to the user, efforts should be made to build a system with more user resilience performance that balances the user's personal interactive experience with the orderliness of the interaction.

## Acknowledgements

Funding: This work was supported by the National Natural Science Foundation of China [grant numbers 61741206].

Tribal Theater Model 18